\documentclass[%
 aip,
 amsmath,amssymb,
 reprint,%
]{revtex4-1}

\usepackage{graphicx}% Include figure files
\usepackage{dcolumn}% Align table columns on decimal point
\usepackage{bm}% bold math
\usepackage[utf8]{inputenc}
\usepackage[T1]{fontenc}
\usepackage{mathptmx}
\usepackage{etoolbox}

\makeatletter
\def\@email#1#2{%
 \endgroup
 \patchcmd{\titleblock@produce}
  {\frontmatter@RRAPformat}
  {\frontmatter@RRAPformat{\produce@RRAP{*#1\href{mailto:#2}{#2}}}\frontmatter@RRAPformat}
  {}{}
}%
\makeatother
\begin{document}

\preprint{AIP/123-QED}

\title{Automated Polarization Rotation for Multi-Axis Rotational-Anisotropy Second Harmonic Generation Experiments}
% Force line breaks with \\
\author{Karna A. Morey*}
\affiliation{Department of Physics, Massachusetts Institute of Technology, Cambridge, MA 02139}
\affiliation{Department of Physics, Stanford University, Stanford, CA 94305}
\author{Bryan T. Fichera*}
\affiliation{Department of Physics, Massachusetts Institute of Technology, Cambridge, MA 02139}
\author{Baiqing Lv}
\affiliation{Tsung-Dao Lee Institute, School of Physics and Astronomy, and Zhangjiang Institute for Advanced Study, Shanghai Jiao Tong University, Shanghai 200240, China}
\affiliation{Department of Physics, Massachusetts Institute of Technology, Cambridge, MA 02139}
\author{Zonqi Shen}
\affiliation{Department of Physics, Massachusetts Institute of Technology, Cambridge, MA 02139}
\author{Nuh Gedik$^\dagger$}
\affiliation{Department of Physics, Massachusetts Institute of Technology, Cambridge, MA 02139}

\date{\today}% It is always \today, today,

\begin{abstract}
Rotational anisotropy second harmonic generation (RA-SHG) is a nonlinear optical technique used to probe the symmetry of condensed matter systems. 
Measuring the dependence of the SHG susceptibility on one or more external parameters, notably strain, field, temperature, or time delay, is an extremely powerful way to probe complex phases of quantum materials. 
Experimentally, extracting maximal information about the SHG susceptibility tensor requires measurements of S and P polarized input and output combinations, which naturally involves the rotation of the polarizers during data collection.  
For multi-axis experiments, this has proved challenging since polarization rotation is typically done manually. 
Automating this process eliminates labor constraints, reduces uncertainty due to low-frequency noise, and expands the type of multi-axis datasets that can be collected; however, it is difficult due to geometrical constraints within the setup.
In this work, we design and implement low-cost, high-fidelity automated polarization rotators for use in multi-axis RA-SHG.
These polarization rotators utilize an electrical slip ring to transfer power to the rotating RA-SHG optical setup as well as a miniature stepper motor to perform the polarization rotation.
We demonstrate this automated system in time-resolved RA-SHG measurements in the non-centrosymmetric semiconductor GaAs. 
For the multi-axis measurements described above, this automated system permits data averaging over longer periods, vastly expedites data collection, and expands the setup measurement capability.
This ultimately opens new frontiers in probing quantum materials using multiple tunable external parameters.
\\\\
* these authors contributed equally to this work\\
${}^\dagger$ corresponding author, gedik$@$mit.edu
\end{abstract}

\maketitle

\section{\label{sec:intro} Introduction}

Probing the structure of crystalline solids is essential to understanding their intrinsic functionalities. 
Traditionally, diffraction techniques based on the scattering of e.g. x-rays\cite{Warren1991X-rayDiffraction} are used to determine this structure; however, these techniques predominantly measure the total electron density and are thus insensitive to long-range ordering of the valence electron subsystem. 
In contrast, nonlinear scattering techniques at optical frequencies like rotational-anisotropy second harmonic generation (RA-SHG) are sensitive rather to the total charge density, and thus offer a complementary view of the electronic, magnetic, and lattice properties of quantum materials \cite{Torchinsky2014ASymmetries, Fichera2020SecondSymmetry, Boyd1992NonlinearOptics, Shen1989OpticalInterfaces, Pisarev2005Second-harmonicReview}. In this paper, we discuss rotational anisotropy second harmonic generation (RA-SHG), although the results are fully generalizable to arbitrary harmonics, e.g. third harmonic generation \cite{Torchinsky2014ASymmetries, Lafrentz2012OpticalEuSe}.

Furthermore, due to the non-invasive nature of RA-SHG, it can also be used to investigate changes along one or more measurement axes, such as strain, field \cite{Lafrentz2012OpticalEuSe}, pressure\cite{Li2022High-pressureTaAs}, temperature\cite{Harter2017ACd2Re2O7, Zhao2017AO, DeLaTorreMirrorCuprate, JinObservationFields,Ahn2024ElectricSemimetal}, or time delay \cite{Luo2021UltrafastGeneration, Shan2021GiantEngineering, Carbin2023EvidenceTransition, Gao2024GiantMultiferroic}.
The phase diagram of quantum materials is heavily influenced by these independent variables, and thus, changes to the SHG response along such measurement axes are of great interest.
Such measurements can illuminate the roles that charge, spin, and thermal degrees of freedom play within quantum materials.

Second harmonic generation gleans microscopic information through the coupling of an order parameter to the second harmonic susceptibility tensor (at least a third-rank tensor with 18 independent components) \cite{Boyd1992NonlinearOptics, Tom1986ObservationSurface, Yang2009SecondCrystals, Pan1989OpticalSurfaces, Dahn1996SymmetryAntiferromagnets, Kirilyuk2005Magnetization-induced-second-harmonicInterfaces}. 
This tensor governs the relationship between the vector properties (i.e. polarization and intensity) of the incident radiation and the generated second harmonic, as shown in Equation \ref{eq:intensity}.
The complexity of this tensor means that a significant amount of data is needed to fully resolve the tensor components, posing experimental challenges with data collection.
The importance of resolving the tensor components can be seen by considering Neumann's principle, which states that these tensors must be invariant under symmetry operations of the material's point group, which constrains the tensors' independent and nonzero elements \cite{Birss1966SymmetryMagnetism}. 
By measuring the intensity and polarization of the generated second harmonic radiation, RA-SHG provides information about these tensor elements and thus the underlying point group.
When performing experiments with more than one measurement axis, the data constraints mentioned above can often be especially burdensome, limiting the type of experiments that can be performed using RA-SHG.
Thus, new methods for collecting data efficiently for multi-axis RA-SHG experiments are incredibly important.

\begin{figure*}
    \centering
    \includegraphics[width = 0.7\textwidth]{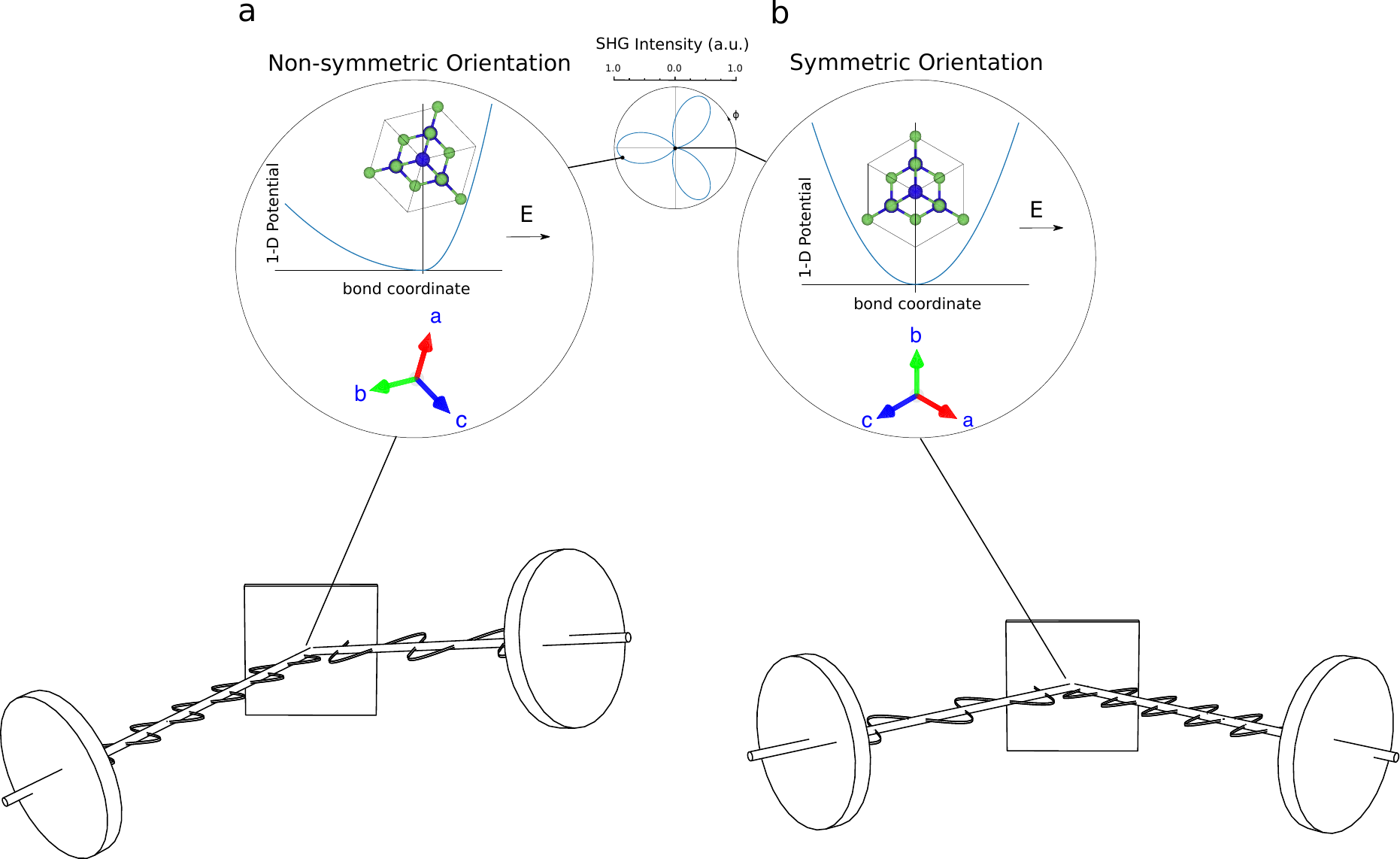}
    \caption{A demonstration of rotational anisotropy second harmonic generation (RA-SHG) in the test sample GaAs. 800 nm light comes in at an oblique angle of incidence, after being passed through a polarizer. The polarization axis of the light is in the plane of incidence and is denoted in the insets of the figure with a black arrow. Depending on the angle of the plane of incidence relative to the crystallographic axes, as well as the polarization of the beam relative to that incoming plane of incidence, the second harmonic response to the stimuli should vary from zero points (nodes) to non-zero maxima.}
    \label{fig:1}
\end{figure*}

A more detailed understanding of the information that needs to be collected to fully resolve the second harmonic susceptibility can be seen by considering figure \ref{fig:1}.
Figure \ref{fig:1} shows a schematic of an RA-SHG setup, where near-infrared light from the laser source enters at oblique incidence and is scattered off the front of the sample, generating blue light at the second harmonic frequency.
To leading order, the measured second harmonic intensity is given by
\begin{equation}
I(\phi) \propto \big| \hat{e}^{\mathrm{out}}_i(\phi) \chi^{(2)eee}_{ijk} \hat{e}^{\mathrm{in}}_j(\phi) \hat{e}^{\mathrm{in}}_k(\phi)\big|^2,
    \label{eq:intensity}
\end{equation}
where $\phi$ is the azimuthal angle between the crystallographic axes and the plane of incidence, $\hat{e}^{\mathrm{out}}$ is a unit vector indicating the direction of the output polarizer, $\hat{e}^{\mathrm{in}}$ is a unit vector indicating the direction of the input polarizer, and $\chi^{(2)eee}$ is the SHG electric-dipole susceptibility tensor \cite{Boyd1992NonlinearOptics, Torchinsky2017}.
For certain values of $\phi$, as shown in figure \ref{fig:1}a, the sample is symmetric with respect to the polarization axis of the incoming light; in this case, the second harmonic generation is constrained to be zero, whereas in general arbitrary angles give a non-zero second-harmonic response (see figure \ref{fig:1}b).
Neumann's principle dictates that equation $\ref{eq:intensity}$ captures the symmetry considerations shown in figure \ref{fig:1}, i.e. the symmetries of the crystal are embedded into its nonlinear susceptibility tensor $\chi^{(2)}$.

These considerations demonstrate the necessity of measuring the full rotational anisotropy of the SHG response, as well as its dependence on the incoming and outgoing polarization directions (which may be P or S-polarized, leading to four independent polarization channels).
The former requires rotating the plane of incidence, which is typically done by rotating the optical setup itself rather than the sample \cite{Petersen2006Nonlinear7,Torchinsky2014ASymmetries, Lu2018FourierGeneration, Fichera2020SecondSymmetry, Fichera2025Light-inducedSemiconductor}.
The geometric constraints involved in rotating the optical setup make it challenging to use electromechanical components to rotate the polarizers shown in figure \ref{fig:1} between S and P configurations, since naively it is difficult to transfer power to the rotating frame of reference.
Because of this, switching between different polarization channels is typically done manually, resulting in time and labor constraints that often limit data collection beyond a single polarization combination \cite{Shan2021GiantEngineering}.
Such constraints severely restrict the ability to infer tensor elements from the RA-SHG data and limit RA-SHG to studies involving often just a single tuning parameter.

Previous multi-axis RA-SHG studies typically either collected exclusively one polarization channel \cite{Shan2021GiantEngineering} (for more than one axis) or simply sacrificed the quantity of collected data and therefore inferred statistics. 
However, these approaches can often miss certain features in RA-SHG data that would be apparent if more efficient data-taking methods were available.
For example, phase transitions in which the order parameter couples differently to different polarization channels through the second harmonic susceptibility \cite{Fichera2025Light-inducedSemiconductor} are best characterized by taking the full rotational anisotropy data.
The benefit of collecting all four polarization combinations underscores the appeal of automated polarization control of RA-SHG setups.

\begin{figure*}
    \centering
    \includegraphics[width = 0.7\textwidth]{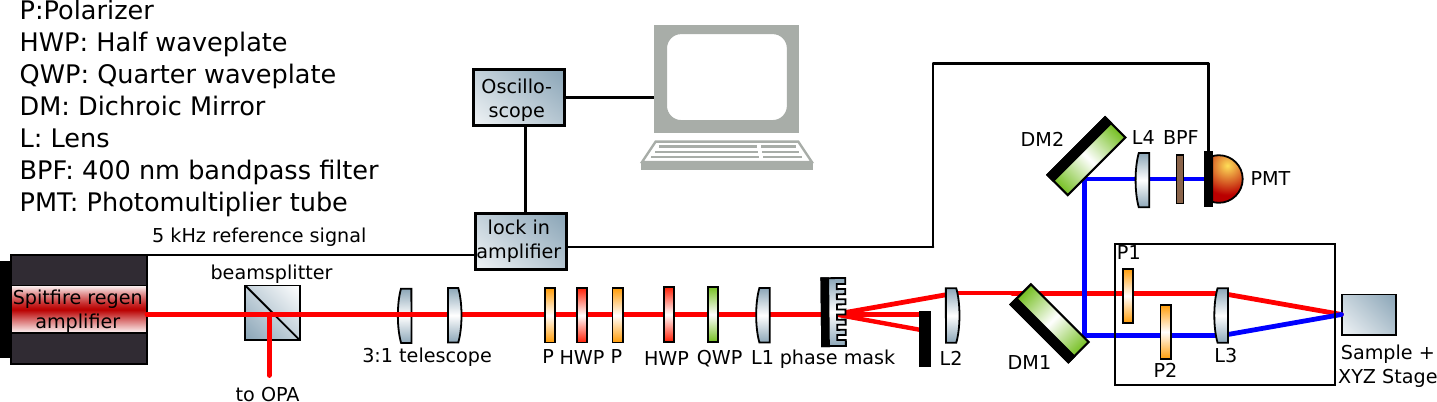}
    \caption{a. A diagram of the full second harmonic generation setup developed and described in \onlinecite{Fichera2020SecondSymmetry}. Boxed in part of the setup shows the area of interest of this work, where the incoming and outgoing polarization channels are set.}
    \label{fig:2}
\end{figure*}

In this work, we design and implement automated polarization rotators for use in multi-axis RA-SHG measurements. 
The devices utilize a miniature stepper motor housed in an electrical slip ring to circumvent the geometrical constraints imposed by the rotating plane of incidence.
In section \ref{sec:disc}, we discuss the specific application of automated polarization rotation to time-resolved RA-SHG and show that it not only expedites data collection but also reduces the setup's sensitivity to low-frequency noise. 
The advantages afforded by the automated polarization rotators exist not only for time-resolved measurements but for any measurements where an external parameter is being varied and compound rapidly as multiple parameters are varied at the same time.

\section{Design}
\label{sec:des}

A detailed schematic of the RA-SHG setup is shown in figure 2.
Telescoped pulses from a 5 kHz pulsed regenerative amplifier (Spectra-Physics Spitfire) are attenuated using a half-waveplate set between two polarizers and then elliptically polarized using a half-waveplate and quarter-waveplate in series. 
The light is then focused by a lens onto a rotating phase mask (which separates the beam into different diffraction orders), and a beam block selects the +1 order beam which is then collimated using another lens (L2).
Finally, the beam passes through a dichroic mirror and a polarizer (P1) before being focused by a third lens (L3) onto the sample.
The reflected radiation is passed through an analyzer (P2) and is redirected using a dichroic mirror periscope, which has equal reflectivity for S and P-polarized light and selects exclusively the 400 nm reflected radiation.
The two polarizers (P1 and P2) and the phase mask are mounted on a rotating shaft coupled to an optical rotary encoder.
After the dichroic mirrors, the second harmonic radiation is focused by a final lens (L4) onto a 400 nm bandpass filter and photomultiplier tube.
The signal from the photomultiplier tube is sent to a lock-in amplifier synced to the 5 kHz pulsed laser reference signal and the amplified signal is then correlated with the signal from the optical rotary encoder using a home-built oscilloscope. 

Importantly, the geometrical constraints introduced by the rotating plane of incidence mean that rotating P1 and P2 between S and P configurations electromechanically is challenging, as power must be transmitted from the stationary laboratory frame to electrical components lying in the rotating frame.

To solve this problem, we utilize a hollow-bore electrical slip ring and a miniature stepper motor, shown in an exploded-view diagram in figure \ref{fig:3}a.
A slip ring uses stationary conductive brushes sliding against a rotating through-bore cylinder to transfer power from a stationary reference frame to a rotating one, as shown in figure \ref{fig:3}b \cite{Argibay2010AsymmetricEnvironment}. 
To accommodate the further requirement that the beams pass through the entire device unimpeded, we employ an 8 mm stepper motor that fits in between the two beams.
Figures \ref{fig:3}c and \ref{fig:3}d show a front view of the device, with the wire grid polarizer in the P and S positions, respectively, and the polarization direction of the polarizer shown in arrows on the edge of the polarizer.
A full rendering, to scale, of the automated system within the full setup is shown in figure \ref{fig:6}. 

\subsection{Slip Ring}

\begin{figure*}
    \centering
    \includegraphics[width = 0.8\textwidth]{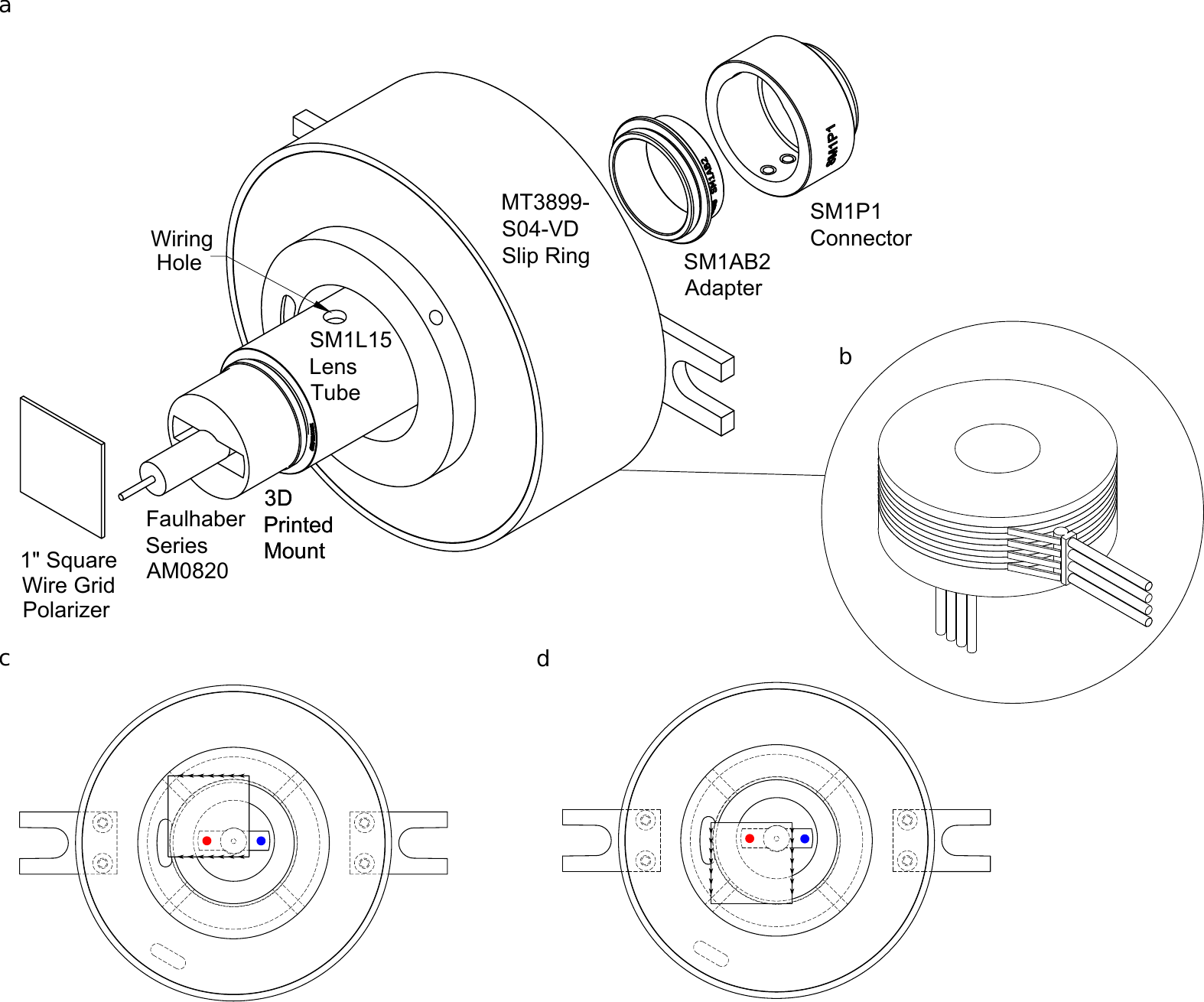}
    \caption{a. An exploded-view diagram of one of the automated polarization rotators, with the individual components labeled. Each of these is a manufactured part except for the 3D-printed mount. b. A schematic of the interior of an electrical slip ring, where low-friction metal brushes allow for electrical conduction across a rotating assembly. c. A front view of the device, in the P-polarized state, with arrows on the edge of the polarizer indicating the polarization direction. d. Same as c, but in the S-polarized state.}
    \label{fig:3}
\end{figure*}

Slip rings are a standard electromechanical device for transferring power from a stationary assembly to a rotating assembly.
An inner, rotating part of the slip ring can freely rotate relative to an outer, stationary part without compromising signal or power transmission.
This functionality is enabled by low-friction metal brushes that slide against another set of electrical contacts, allowing free rotation while maintaining good electrical conductivity \cite{Argibay2010AsymmetricEnvironment}, as shown in figure \ref{fig:3}b.
To allow for the uninterrupted passage of the beams through the device, we use a hollow-bore slip ring (Moflon MT3899-S040-VD) whose 38mm inner bore rotates along with the lens tube-pulley assembly shown in figure \ref{fig:6}.
Either set screws located on the front side of the slip ring (as shown in figure \ref{fig:3}) or a 3D-printed cylindrical hollow bore adapter (not shown) can secure the inner bore of the slip ring to the lens tube.

\subsection{Miniature Stepper Motor}
\label{sec:stepper}
The hollow bore slip ring allows for the transfer of power to the rotating shaft.
To rotate the wire grid polarizer between S and P configurations while also allowing both beams to pass through the entire device, we use an 8 mm diameter Faulhaber Series AM0820 stepper motor.
The stepper motor's shaft is epoxied to a 1-inch square wire-grid polarizer roughly 5 mm from the closest horizontal and vertical edges, as shown in figures \ref{fig:3}c and d.
The stepper motor is mounted in the center of the SM1L15 lens tube using a custom, 3D printed mount as shown in figure \ref{fig:3}a. 
This mount, along with careful epoxying of the polarizer to the motor shaft, ensures that the incidence angle of the beam is close to normal.
Furthermore, the mount ensures that the stepper motor's shaft is aligned closely to the central rotation axis of the entire assembly, minimizing torques on the motor and polarizer.
This stepper motor can support up to 800 total micro-steps per revolution
(0.45$^\circ$ per micro-step) and 0.65 mNm of holding torque, and is small enough to fit in between the incoming and outgoing beams. 

The motor can then move the polarizer between the P configuration (figure \ref{fig:3}c) and the S configuration (figure \ref{fig:3}d) without manual intervention or the need to stop the rotation of the entire assembly.
The offset placement of the polarizer on the motor shaft allows only one of the beams (either incoming or outgoing) to pass through the polarizer.
The SM1AB2 and SM1P1 adapter pieces allow for arbitrary rotation relative to the longer lens tube-pulley assembly, allowing the beams to be aligned with the empty spaces in the stepper motor mount.
These adapter pieces also allow for small (+/- 20deg) rotations of the polarizer about the central axis to account for factory defects which may cause the polarization axis to be misaligned with the straight edges of the polarizer.

\begin{figure*}
    \centering
    \includegraphics[width = 0.8\textwidth]{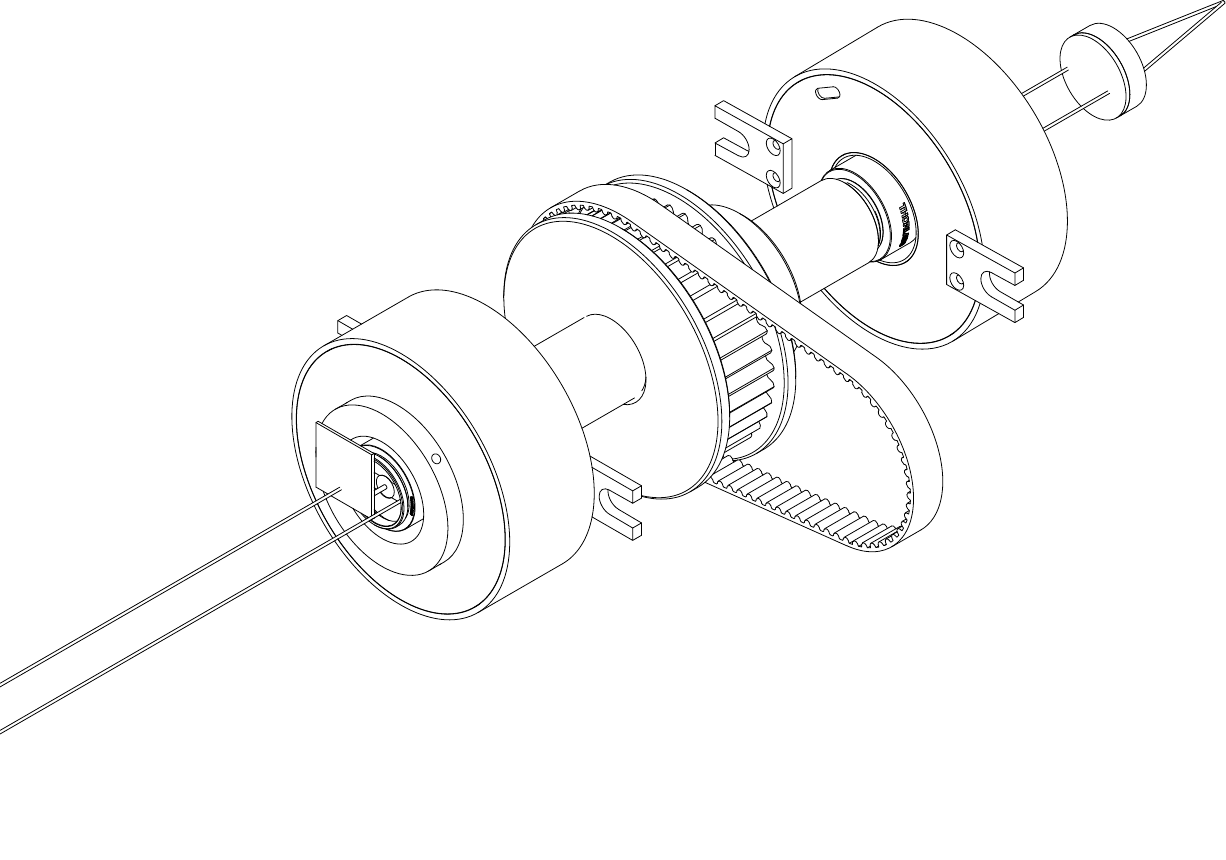}
    \caption{A 3-dimensional rendering of the boxed-in part of the setup in figure \ref{fig:2}, with fully automated polarization rotators.}
    \label{fig:6}
\end{figure*}

Another copy of the same system shown in figure \ref{fig:3} is mounted on the other side of the lens tube-pulley assembly and controls the polarization of the reflected beam, as shown in figure \ref{fig:6}. 
The back sides of both devices, shown in figure \ref{fig:6}, are attached to the rotating lens tube-pulley-motor-shaft assembly shown in figure \ref{fig:6} via threads on the long central lens tube. 
A hole drilled in the top and bottom of the SM1L15 piece in each device shown in figure \ref{fig:3}a allows feedthrough wires to connect the four leads of the stepper motor to the four rotating inner-bore slip ring leads.
The four leads on the stationary side of each slip ring directly connect to a Geckodrive G250X digital step driver for each device, controlled by a single Arduino Nano Every.
This control circuit allows for the integration of the stepper motors with existing instrumentation software.

Whenever the stepper motors lose power, the polarizers need to be re-aligned due to the loss of their holding torque.
The alignment is performed by using a reference polarizer with a known polarization axis.
First, the plane of incidence of the entire assembly is positioned to the known polarization plane of the reference polarizer, then the polarizers themselves can be oriented by placing the beam through the reference polarizer and rotating the stepper motor until beam extinction (yielding an S-polarized alignment).  
The polarizers can then freely rotate between this S-polarized position and the orthogonal P-polarized position.

\section{Validation and Discussion}
\label{sec:disc}
\begin{figure*}
    \centering
    \includegraphics[width = 0.8\textwidth]{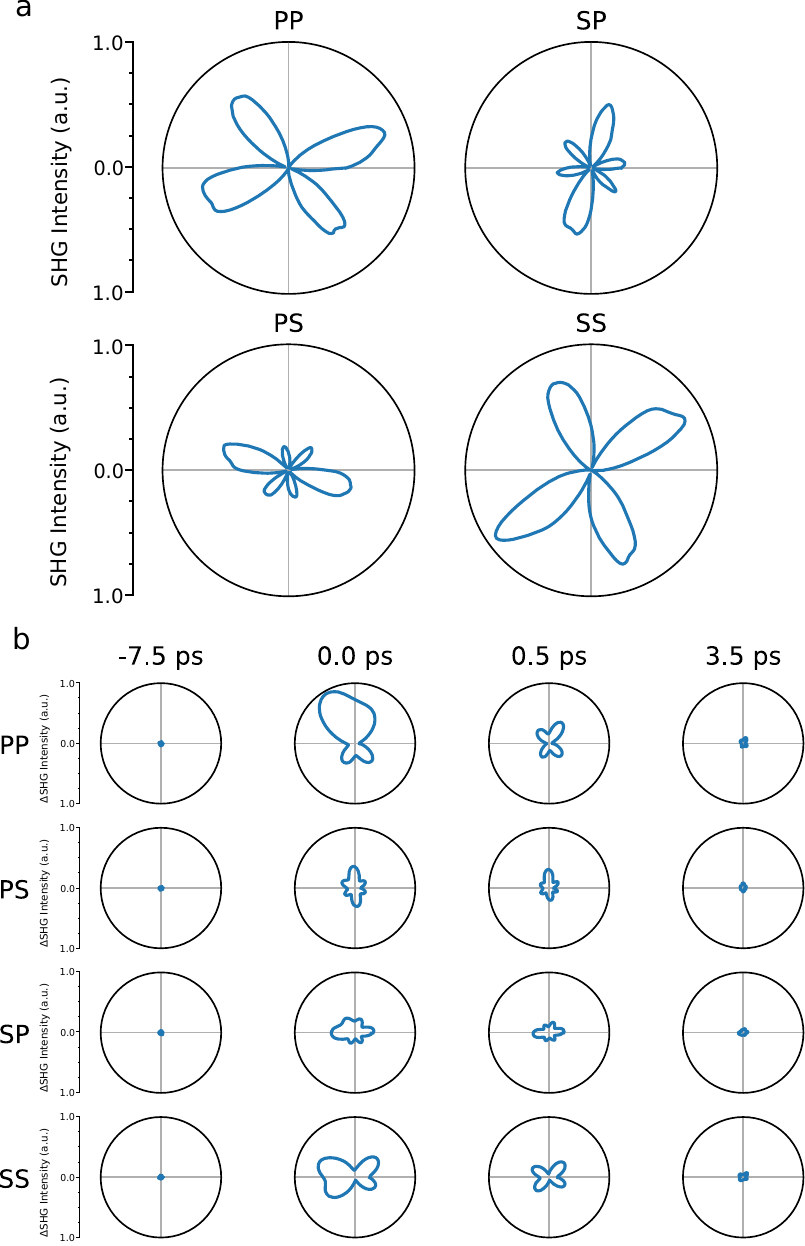}
    \caption{a. A polar plot of the equilibrium SHG intensity for the test sample GaAs b. Change in SHG intensity (from the pre-pump state) as a function of time since zero-delay for the test sample GaAs. This is a demonstration of using the automated polarization rotators to take time-resolved SHG data. 55 different pump-probe delays (only four time delays are plotted) were taken in this dataset, meaning that 220 polarization rotations were performed using the automated polarizers.}
    \label{fig:4}
\end{figure*}

We note that in our experience, the polarization rotators display a high degree of angular fidelity, with an alignment error of less than 0.5 degrees over 5000 rotations of the stepper motors. 
To test these automated polarization rotators, we performed time-resolved RA-SHG on the non-centrosymmetric semiconductor GaAs \cite{Ducuing1963ObservationCrystals, Torchinsky2014ASymmetries, Niu2015High-speedDetection}. Figure \ref{fig:4}a shows the equilibrium SHG intensity for the four polarization states. 
For these equilibrium measurements, we use a lock-in amplifier at a frequency of 5 kHz, or the full repetition rate of the laser, since there is no pump line. 
To measure time-resolved RA-SHG in GaAs, a pump line was added to the setup shown in figure \ref{fig:2}, by mounting a 45$^\circ$ mirror onto L3 so that the pump beam may be reflected directly onto the sample. 
The pump pulse is produced by an optical parametric amplifier (Spectra-Physics Topas) tuned to a wavelength of 1.3 $\mu$m, and was circularly polarized.
The pump line includes a delay stage, which controls the relative time delay between the pump and the probe pulses, and a chopper wheel, which reduces the effective repetition rate of the pump pulses to 2.5 kHz. 
With the pump repetition rate of 2.5 kHz, and the probe repetition rate of 5 kHz, locking in at a frequency of 2.5 kHz produces a signal proportional to the change in the SHG intensity relative to equilibrium.
For time-resolved measurements on GaAs, shown in figure \ref{fig:4}b, 55 different pump-probe delays were taken, resulting in 220 different automatic rotations of the polarizers.
Because of the different methods for lock-in detection between the equilibrium and time-resolved data, we note that there is an unknown scale factor between the arbitrary units used in the respective measurements.
Calibration of such scale factor between the two techniques is beyond the scope of this work, but studies such as \cite{Sirica2018LightinducedTS}, whose setup is very similar to ours, demonstrate the relative sensitivity of this technique in detecting changes in the SHG intensity relative to equilibrium.
The data in figure $\ref{fig:4}$b shows the change in the SHG intensity relative to equilibrium as a function of the pump-probe delay, for each polarization channel.
We note that the asymmetric nature of the polar plots in figure \ref{fig:4}b is due to a slight mismatch in spatial overlap between pump and probe pulses, which time-resolved RA-SHG in GaAs is very sensitive too. 
However, the purpose of these time-resolved measurements is mainly to demonstrate the stability and fidelity of the polarization rotators, rather than a detailed quantitative analysis of the electron dynamics within GaAs.
The use of the polarization rotators in this validation measurement demonstrates their functionality and broad utility for multi-axis RA-SHG measurements.

This system was also used to expedite data collection in recent studies of time-resolved, temperature-dependent RA-SHG \cite{Fichera2025Light-inducedSemiconductor}.
This particular type of RA-SHG measurement demonstrates the general advantages of automated polarization control in multi-axis RA-SHG measurements.
To see this, note that if the rotation of the polarizers is performed manually, the best data-taking strategy involves taking the time dependence of one polarization channel in full before moving to the next. 
However, in the presence of low-frequency noise (due to e.g. fluctuations in the laser intensity), this is problematic as the different polarization channels will not be taken under equivalent conditions. 
In the automated case, in contrast, all polarization channels can be taken at each time delay. 
This greatly decreases the experiment's susceptibility to low-frequency noise and expedites data taking by allowing for overnight and multi-day scans.

Automated polarization rotation thus vastly improves RA-SHG data collection, eliminating the need for manual polarization rotation, especially in cases of multiple experimental axes. 
The devices are inexpensive and easily implemented in existing RA-SHG setups.
Multi-axis measurements are becoming increasingly important to resolve complicated phase diagrams and competing orders, and thus future measurements will increasingly rely on automated systems such as the one presented in this work.

\begin{acknowledgments}
K.M. would like to acknowledge the generous support of the MIT UROP program during this work. K.M., B.F., B.L., Z.S., and N.G. acknowledge support from the US Department of Energy, BES DMSE (data taking and analysis), and Gordon and Betty Moore Foundation’s EPiQS Initiative grant GBMF9459 (instrumentation).
\end{acknowledgments}

\section*{Data Availability Statement}

The data that support the findings of this study are available from the corresponding author upon reasonable request.

\section{References}
\bibliography{references}% Produces the bibliography via BibTeX.

\end{document}